\documentclass[twoside,twocolumn]{article}
\pdfoutput=1
\pdfoutput=1 % if your are submitting a pdflatex (i.e. if you have
\usepackage{amssymb}
\usepackage{graphicx}
\usepackage{lineno}
\usepackage{subfigure}
\usepackage{amsmath}
\usepackage{graphicx}
\usepackage{hyperref}
\hypersetup{colorlinks=true,urlcolor=blue}
\DeclareMathOperator*{\argmin}{argmin}

\usepackage[symbol]{footmisc}

\usepackage{tabulary,graphicx,times,caption,fancyhdr,amsmath}
\usepackage[utf8]{inputenc}
\usepackage[numbers,sort&compress]{natbib}
\usepackage[paperheight=11in,paperwidth=8.3in,margin=2.5cm,headsep=.7cm,top=2.7cm]{geometry}
 \date{} \emergencystretch 8pt
\usepackage{url,multirow,morefloats,floatflt,cancel,textcomp} %,tfrupee}
\usepackage{pifont}
\usepackage[nointegrals]{wasysym}
\makeatletter

\AtBeginDocument{
\expandafter\ifx\csname eqalign\endcsname\relax
\def\eqalign#1{\null\vcenter{\def\\{\cr}\openup\jot\m@th
  \ialign{\strut$\displaystyle{##}$\hfil&$\displaystyle{{}##}$\hfil
      \crcr#1\crcr}}\,}
\fi
}

\let\lt=<
\let\gt=>
\def\processVert{\ifmmode|\else\textbar\fi}

\@ifundefined{subparagraph}{
\def\subparagraph{\@startsection{paragraph}{5}{2\parindent}{0ex plus 0.1ex minus 0.1ex}%
{0ex}{\normalfont\small\itshape}}%
}{}

\newcommand\role[1]{\unskip}
\newcommand\aucollab[1]{\unskip}
  
\@ifundefined{tsGraphicsScaleX}{\gdef\tsGraphicsScaleX{1}}{}
\@ifundefined{tsGraphicsScaleY}{\gdef\tsGraphicsScaleY{.9}}{}
\def\checkGraphicsWidth{\ifdim\Gin@nat@width>\textwidth
	\tsGraphicsScaleX\textwidth\else\Gin@nat@width\fi}

\def\checkGraphicsHeight{\ifdim\Gin@nat@height>.9\textheight
	\tsGraphicsScaleY\textheight\else\Gin@nat@height\fi}

\def\fixFloatSize#1{\@ifundefined{processdelayedfloats}{\setbox0=\hbox{\includegraphics{#1}}\ifnum\wd0<\columnwidth\relax\renewenvironment{figure*}{\begin{figure}}{\end{figure}}\fi}{}}
\let\ts@includegraphics\includegraphics

\def\inlinegraphic[#1]#2{{\edef\@tempa{#1}\edef\baseline@shift{\ifx\@tempa\@empty0\else#1\fi}\edef\tempZ{\the\numexpr(\numexpr(\baseline@shift*\f@size/100))}\protect\raisebox{\tempZ pt}{\ts@includegraphics{#2}}}}

\AtBeginDocument{\def\includegraphics{\@ifnextchar[{\ts@includegraphics}{\ts@includegraphics[width=\checkGraphicsWidth,height=\checkGraphicsHeight,keepaspectratio]}}}

\def\URL#1#2{\@ifundefined{href}{#2}{\href{#1}{#2}}}

\def\UrlOrds{\do\*\do\-\do\~\do\'\do\"\do\-}%
\g@addto@macro{\UrlBreaks}{\UrlOrds}
\makeatother

\makeatletter\def\hindawiIndent{3pc}
\def\author#1{\gdef\@author{\hskip-\dimexpr(\tabcolsep)\hskip\hindawiIndent\parbox{\dimexpr\textwidth-\hindawiIndent}{\raggedright\bfseries#1}}}
\def\title#1{\gdef\@title{\raggedright\bfseries\ifx\@articleType\@empty\else\@articleType\\\fi#1}}
\let\@articleType\@empty \def\articletype#1{\gdef\@articleType{{\normalfont\itshape#1}}}
\let\@runningHead\@empty \def\RunningHead#1{\gdef\@runningHead{{\normalfont #1}}}
\fancypagestyle{headings}{\fancyhf{}\fancyhead[C]{\ifx\@runningHead\@empty\else\@runningHead\fi}\fancyhead[R]{\thepage}}
\makeatother

\begin{document}
\twocolumn[
\begin{@twocolumnfalse}
\title{Deep Learning the Effects of Photon Sensors on the Event Reconstruction Performance in an Antineutrino Detector}
\author{Chang-Wei Loh\textsuperscript{1},
	Zhi-Qiang Qian\textsuperscript{1}\thanks{Coauthor},
	Rui Zhang\textsuperscript{1},
	You-Hang Liu\textsuperscript{1},
	De-Wen Cao\textsuperscript{1},
	Wei Wang\textsuperscript{1},
	Hai-Bo Yang\textsuperscript{1} and
            Ming Qi\textsuperscript{1}\thanks{Corresponding author}
            ~\\[-3pt]\normalsize\normalfont\itshape 
~\\\textsuperscript{1}{Physics Department\unskip, Nanjing University, 22 Hankou Road, Nanjing, Jiangsu, China}}

\maketitle 

%\begin{abstract}
We provide a fast approach incorporating the usage of deep learning for studying the effects of the number of photon sensors in an antineutrino detector on the event reconstruction performance therein. This work is a first attempt to harness the power of deep learning for detector designing and upgrade planning. Using the Daya Bay detector as a case study and the vertex reconstruction performance as the objective for the deep neural network, we find that the photomultiplier tubes (PMTs) at Daya Bay have different relative importance to the vertex reconstruction. More importantly, the vertex position resolutions for the Daya Bay detector follow approximately a multi-exponential relationship with respect to the number of PMTs and hence, the coverage. This could also assist in deciding on the merits of installing additional PMTs for future detector plans. The approach could easily be used with other objectives in place of vertex reconstruction.

%\end{abstract}
\end{@twocolumnfalse} \vspace{0.6cm}
] \footnote[1]{First co-author}\footnote[2]{Corresponding author}

\section*{}
\vspace{-1cm}
    
\section{Introduction}
\label{sec:intro}
The choice of photon sensors such as photomultiplier tubes (PMTs), be it their expected sizes, locations and the total number of sensors in antineutrino detectors, including Daya Bay \cite{CAO201662}, Double Chooz \cite{Chooz}, RENO  \cite{RENO} and JUNO \cite{JUNO} are of interest as these sensors are the information gatherers through which we can identify antineutrino interaction events. This work is an attempt in using machine learning, in particular deep learning \cite{LeCun:2015a, 10.1007/978-3-642-39593-2_1, SCHMIDHUBER201585} as a way to understand how the number of PMTs in the detector influences the event reconstruction performance, and extract lessons to be learned therefrom for areas such as detector designing and upgrade planning. To the best of our knowledge, this work is the first study on the efficacy of deep learning in detector designing and planning. For this work, we ask the following: suppose we are given $N$ possible number of locations for the installation of $k$ number of PMTs $(k \leq N)$, where should the PMTs be installed such that the event interaction vertex reconstruction is optimal or near-optimal given only these $k$ PMTs in the detector and $N$ possible locations? Of course, $N$ and $k$ could be infinite, but this is technically impossible as it would not meet the budget of a detector construction. In this work, we use deep learning on a model of the Daya Bay detector as a case study to understand the impact of PMTs on event position vertex reconstructions in a detector. The vertex is useful for studies on signal-background discriminations and the correction to the position-dependent energy response in the detector. The reconstruction of the vertices has been studied in-depth in the Daya Bay experiment using non-machine learning methods. As such, this allows us to cross-check our vertex reconstruction with deep learning with other methods in the Daya Bay before studying its potential in detector designing. Moreover, the vertex reconstruction performance is chosen as the objective since it is relatively simple for deep learning to handle for a clear understanding of the approach without involving too much experimental details.
Nonetheless, experimentalists can easily substitute the vertex reconstruction performance with other objectives of interest. Beyond antineutrino detectors, sensor placements have been studied in areas ranging from water network distributions \cite{Andreas:2008} to fault detections \cite{faultdetection:2000}. 

The Daya Bay antineutrino detectors are
liquid scintillator detectors with a physics program focusing on the precision measurement of the neutrino mixing angle $\theta_{13}$ with reactor antineutrinos. Each Daya Bay detector consists of three concentric cylindrical tanks: an inner acrylic vessel (IAV) containing gadolinium (Gd)-doped liquid scintillator, an outer acrylic vessel (OAV) containing undoped liquid scintillator which surrounds the IAV, and a stainless steel vessel (SSV) which surrounds the IAV and OAV. With this design, the detectors could detect the interaction of the antineutrinos and the scintillator via inverse beta decay (IBD) reactions: 
\begin{equation}
\bar{\nu}_e + p \to e^+ + n.
\end{equation}
The emitted positron then undergoes ionization processes in the liquid scintillator before annihilating with an electron producing a prompt signal with an energy deposition in the range of 1 - 8 MeV. The deposited energy is converted to scintillation photons which are then collected by the PMTs.  As the positron displacement prior to the annihilation is negligible, the interaction vertex of the prompt signal can be assumed to be the antineutrino IBD interaction vertex.
However, the neutron thermalizes and diffuses before being captured on either a proton or Gd with a mean capture time of $\sim30$ $\mu s$ in the Gd-doped liquid scintillator and $\sim200$ $\mu s$ in the undoped liquid scintillator, giving rise to a delayed signal.
A total of 192 Hamamatsu R5912 8-inch PMTs \cite{PMT_waveform}, arranged in a layout with 8 rows and 24 columns, are installed on the vertical wall of the SSV pointing inward towards the OAV and IAV forming a total of 6\% photodetector coverage. 
Located above and below the OAV are reflective panels that serve to redirect scintillation light towards the PMTs thereby increasing the photon collection efficiency to 12\% effectively.

As afore-mentioned, we used deep learning to perform the IBD vertex reconstruction in order to study the effects of PMTs on an event reconstruction. Deep learning is a class of machine learning, which is especially adept at leveraging large datasets to compute human-comprehensible quantities by learning the various degrees of correlations within. Notably, it can, on its own, learn to discover functional relationships from the data without \textit{a priori} given, effectively forming a mapping from the inputs to a quantity of interest. In other words, deep learning seeks to model the quantity of interest $y$ using a vector of inputs $x$ with $DL(x,p) = y$, where $p$ are parameters of the deep network; their numerical values found by minimizing the error between the predicted $y'$ and $y$. 
Deep learning machine architectures, commonly known as deep neural networks (DNN), are based on artificial neural networks~\cite{ANN} but deeper in terms of the number of hidden layers, and are more flexible in terms of how each neuron is connected to other neurons.  

The ubiquity of deep learning and its significant success over traditional methods across disparate fields~\cite{Hinton:2012,Go:2016,Stemcell} in discovering patterns is suprising. However, this may well be due, in part, to that our universe operates on simple physical properties \cite{arxiv:1608.08225}. 
%Neural networks have also been used to explore, among others, the possibility of estimating the lifetime of deep submicron MOSFETs~\cite{0268-1242-20-2-010}, predicting and optimizing the magnetic properties of iron-based materials~\cite{HAMZAOUI200917}, predicting the characteristics of in-flight particles in a nonlinear dynamic system~\cite{Guessasma2004}. 
In high-energy physics, deep learning has demonstrated its prospective use in jets~\cite{Guest:2016iqz,color:2017}, as part of the signal-background discrimination toolkit in the search for beyond the Standard Model particles~\cite{exoticbaldi} and Higgs bosons~\cite{taubaldi}, and in neutrino physics experiments~\cite{Kohn,miniboone,nova,next,neutrinoRacah}.

\section{Recursive Search}
\label{section:recursive}
As mentioned in Section \ref{sec:intro}, we wish to search for the $k$ most important locations corresponding to $k$ PMTs installed therein from the $N$ total possible locations in the detector in determining the vertex position $V$ of events collected from the detector. $N$ is a free parameter which could be chosen during the detector design and simulation stage. Denoting the set containing the $k$ number of most important locations as the set $S^*_{k}$, this implies that we should find the set $S^*_{k}$ such that the vertex reconstruction error is minimized.
However, finding such $k$ locations simultaneously is a task confounded by a computation that grows exponentially with $k$. Alternatively, we could search for an approximation to $S^*_{k}$ by recursively finding the important PMT location one at a time, which can be achieved using deep learning. Since searching for the most important location is equivalent to searching for the most important PMT at that particular location, 
the phrase "$k$-th important PMT" will be used in this work as a shorthand for "$k$-th important PMT location".

Let the true position of the IBD prompt events be $V_{true} = \{x_{true},y_{true},z_{true}\}$; the predicted position using DNN as
$V_{pred} = \{x_{pred},y_{pred},z_{pred}\}$, then in a recursive search, the $k$-th importnat PMT, $PMT_{k}^*$ will be the one that maximizes the improvement in the resolution $\sigma$ of the residual distribution conditioned on already known the $(k-1)$ other PMTs found through the recursive search, i.e., $\{PMT_{k-1}^*,PMT_{k-2}^*,...,PMT_1^*\}$, and where the residual is $V_{pred} - V_{true}$. Namely,
\begin{equation}
%\label{eqn:infogain}
PMT_k^* = \mathop{\argmin}_{P_k \in N \setminus S^{recu}_{k-1}} \sigma(PMT_{k},S^{recu}_{k-1}), %- \sigma(S_{k-1}) \label{eqn1:infogain} \\
%\mathop{\argmin}_{P_k \in N \setminus S_{k-1}} \sigma \\
%&=
%&=\mathop{\argmin}_{P_k \in A \setminus S_{k-1}} h(V | P_{k},S_{k-1}) - h(V | S_{k-1}) \label{eqn1:infogain} \\
%&=\mathop{\argmin}_{P_k \in A \setminus S_{k-1}} h(P_{k} | V, S_{k-1}) - h(P_{k} | S_{k-1}) \label{eqn2:infogain},
\label{eqn:optimize}
\end{equation}
where $S^{recu}_{k-1} = \{PMT^*_{k-1},...,PMT^*_1\}$ and $N$ is the set containing all the PMTs.
Using Equation \ref{eqn:optimize}, PMTs could be progressively added into a larger and larger subset $S$ defining the best set found by the algorithm. Alternatively, one could perform a backward elimination: starting from the set with all PMTs, and progressively eliminating the most "unimportant'' PMT.
At the conclusion of this recursive search, we obtain a curve of the event reconstruction resolution vs. the number of PMTs used for the reconstruction thereof.

\section{Deep Neural Network}
\label{section:DNN}

In our approach utilizing DNNs, we used a Monte Carlo dataset comprising 2 million IBD prompt events obtained from a Daya Bay detector model which were randomly partitioned into a training set (1.4 million), a validation set (0.3 million) and a test set (0.3 million). The validation set is used for the early stopping of the DNN training to prevent overfitting or underfitting of the data \cite{earlystopping}. The parameter $N$ as defined in Section \ref{section:recursive} would be 192 corresponding to the 192 PMT locations in the Daya Bay detector model. The charge information of the 192 PMTs are fed into the DNN as its inputs, and the output is the predicted vertex location $V_{pred}$.
To train the DNN, we used the mean square error ($MSE$) loss function to measure the error between the predicted and the truth vertex positions:
\begin{equation}
%MSE = \frac{1}{T} \sum_{i=x,y,z} \sum_{j=1}^{T} (v_{{ij}_{pred}} - v_{{ij}_{true}})^2 ,
MSE = \frac{1}{T} \sum_{j=1}^{T} (v_{{ij}_{pred}} - v_{{ij}_{true}})^2 ,
\end{equation}
where $T$ is the number of events, $v_{{ij}_{pred}}$ and $v_{{ij}_{true}}$ are the predicted and truth values for the i-th coordinate of the j-th event vertex respectively ($i = x,y,z$). The $MSE$ was minimized to obtain the optimal DNN parameters. The minimization is typically done with a gradient descent method \cite{backprop} involving the gradient of the loss function with respect to the DNN parameters, including the weights of each neuron, i.e. at each training iteration, the parameters $w$ are updated via
\begin{equation}
w_i = w_i - \gamma \cdot \frac{\partial MSE}{\partial w_i},
\end{equation}
where $\gamma$ is the learning rate determined by the user that controls the step length in the negative gradient direction during the training stage. When the $MSE$ reaches a minima, $\frac{\partial MSE}{\partial w_i} = 0$. At this point, the DNN has found the needed parameter values to best reconstruct the vertex position. To train the DNN, the $\gamma$ starts with a value of 0.001 and is progressively multiplied a factor of 0.5 whenever the value of the loss function metric stops improving. In this manner, the DNN training will descent quickly in the direction of the minima in the early stage; with a smaller learning rate at a later stage, the training will not overshoot the minima but will descent steadily towards it. An early stopping is made during the training, whereby the training is terminated when no further improvements could be observed from the loss function value after a pre-determined number of training rounds, in this case ten. Without such early stopping, the loss function value can rise again indicating that an overfitting has occur.

%A representative image of the $8 \times 24$ charge pattern due to an electron-positron annihilation event in the Daya Bay detector is shown in Figure \ref{chargedistribution}. The color of each rectangle corresponds to the charge collected by the PMT at the corresponding ring and column location. 
The efficacy of deep learning to predict the position of the IBD prompt events can be demonstrated by the residual distributions shown in Figure \ref{residual_192sensor} where the charge information of the 192 PMTs are fed into a DNN as its inputs.
The DNN used here to obtain $V_{pred}$ consists of multiple fully-connected layers with ReLU \cite{icml2010_NairH10} hidden neurons. The optimal number of layers and neurons were obtained using a tree-structured Parzen estimator \cite{Bergstra}. The resulting network comprises of three hidden layers containing 180, 148 and 148 neurons respectively.
The resolutions as obtained from the Gaussian fit to the residual distributions are 67 mm and 80 mm for $(x_{pred}-x_{true})$ and $(z_{pred} - z_{true})$ respectively.

%\begin{figure}[!htbp]
%	\centering
%	\includegraphics[trim=100 0 0 0,clip,width=.6\textwidth]{chargedistribution_prompt161_v2}
%	\caption{Charge distribution of a simulated event in the Daya Bay detector as collected by the 192 PMTs. The color scale shows the charge magnitudes. Charge information like these are fed into the DNN as inputs.}
%	\label{chargedistribution}
%\end{figure}

\begin{figure}[!htbp]
	\centering
	\subfigure[$x$] {
		\includegraphics[trim=10 0 20 0, clip, width=0.46\textwidth]{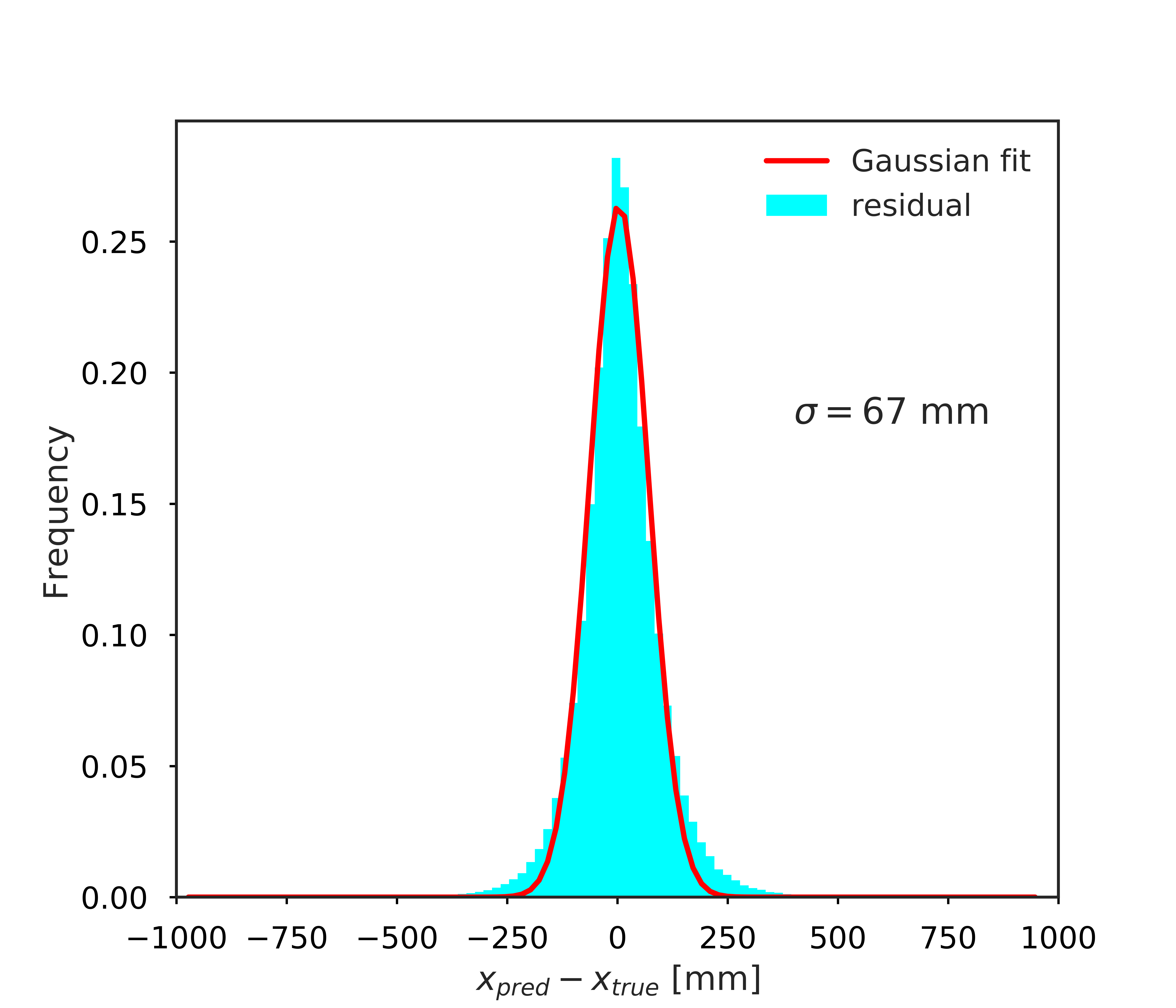}
		\label{fig_firstsub}
	}
	%\subfigure[$y$] {
	%  \includegraphics[width=0.46\textwidth]{residual_Y_192}
	%  \label{fig_secondsub}
	%}
	\subfigure[$z$] {
		\includegraphics[trim=10 0 20 0,clip,width=0.46\textwidth]{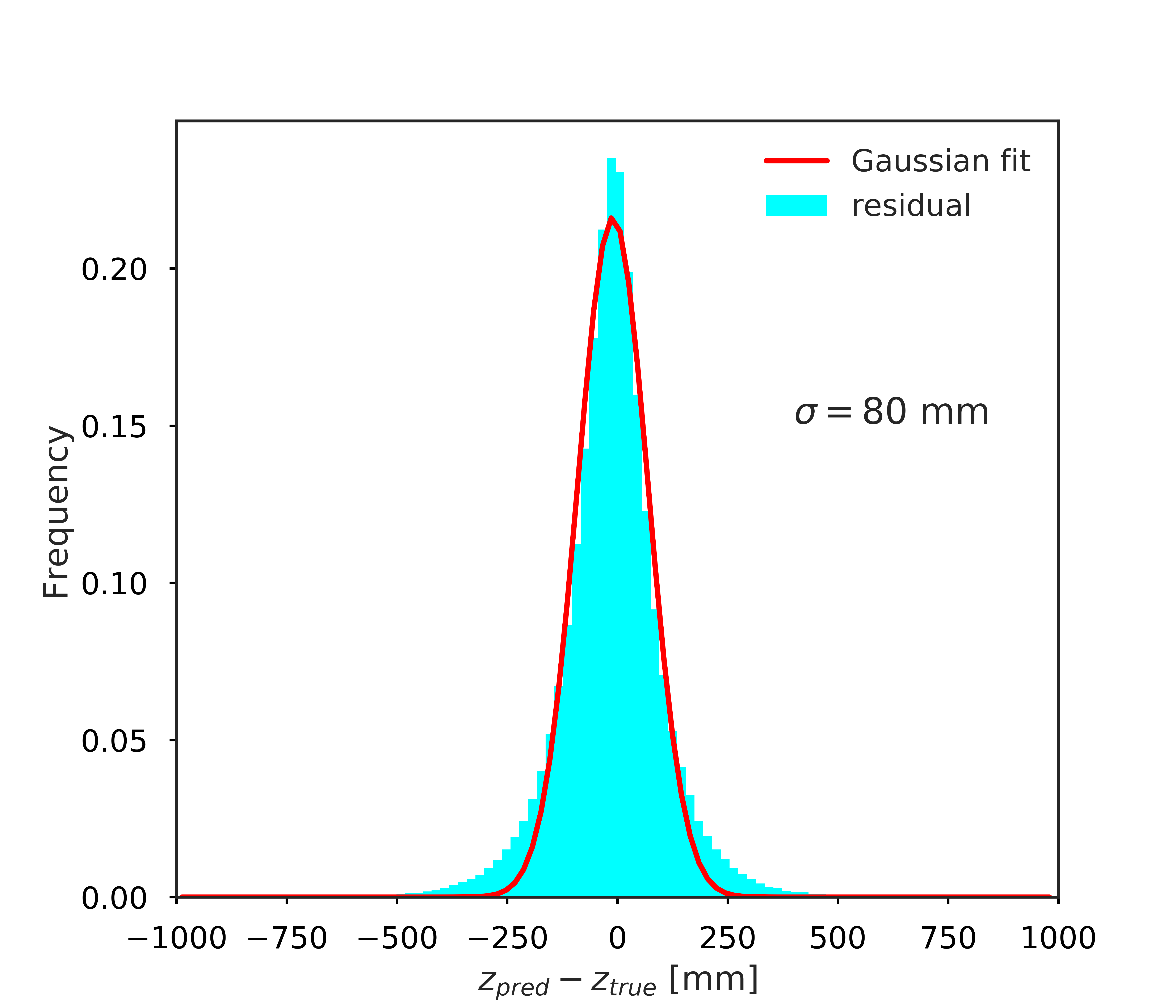}
		\label{fig_thirdsub}
	}
	\caption{Residual distributions for x and z using all 192 PMT charge information.}
	\label{residual_192sensor}
\end{figure}

A straightfoward and brute force use of Equation \ref{eqn:optimize} in a recursive search using a DNN to identify $PMT_k^*$ would be to check over all the remaining PMTs not in the optimal set $S^{*}_{k-1}$ and separately construct the residual distributions; picking the one giving the best resolution for a particular coordinate in $V_{true}$.
For this brute force search, we used a DNN architecture similar to the aforementioned DNN.
The input layer will contain neurons with charge information from the already-chosen PMTs, i.e., those in $S^{recu}_{k-1}$, plus a candidate PMT, i.e., $PMT_k$. The computation time for such a search grows quadratically with the total number of PMTs in the detector. Such a brute force search is clearly not scalable. Hence, in this work, we have also used a fast approach to approximate the brute force search but which mitigates the non-scalability of the latter.

\begin{figure}[!htbp]
	\centering\includegraphics[width=.5\textwidth]{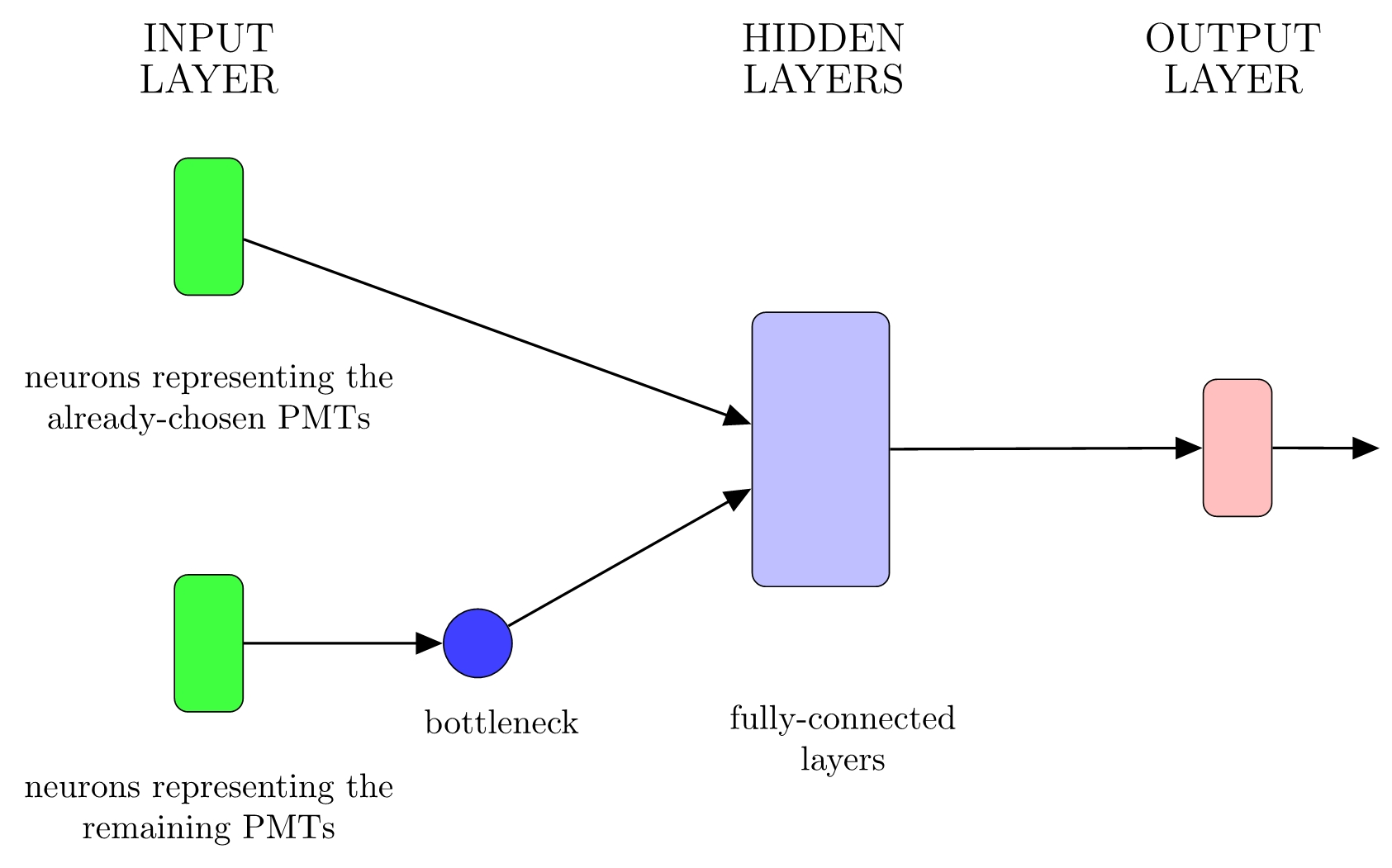} %bottleneck_diagram.png}
	\caption{DNN architecture consisting of a bottleneck neuron.}
	\label{bottleneck_diag}
\end{figure}

This fast approach integrates a DNN component from the autoencoder architecture \cite{NIPS1993_798}: a bottleneck layer with a single neuron, as shown in Figure \ref{bottleneck_diag}.
In this bottleneck DNN architecture, the remaining candidate PMTs not in $S^{recu}_{k-1}$ are forced to connect to the bottleneck neuron before being given to the fully-connected layers as inputs, effectively demanding the DNN to search for the best weights associated with each of these PMTs. 
At the bottleneck region, the DNN computes the sum $\sum w_i PMT_i$,
where $i$ runs over the candidate PMTs, the quantity $PMT_i$ is the $i$-th PMT input to the DNN in the form of charge information and the weight $w_i$ of the $i$-th PMT is a parameter in the DNN. When the training stage of the DNN ends, the $w_i$s would have reached their best values corresponding to a minima of the $MSE$. The PMT with the largest weight in magnitude indicates that the reconstruction of the position of the IBD prompt events relies the heaviest on this PMT compared to the rest of the candidate PMTs. Hence, this PMT would be our $k$-th important PMT, $PMT_k^*$. Crucially, this type of DNN only needs to be trained once to identify $PMT^{*}_k)$ no matter the value of $k$, whereas the brute force DNN needs to be trained $(N-k)$ times, once for each remaining candidate PMT.

\section{Results}

\begin{figure}[!htbp]
	\centering
	\subfigure[$\sigma_x(PMT_1)$] {
		\includegraphics[trim=110 0 110 0,clip,width=0.46\textwidth]{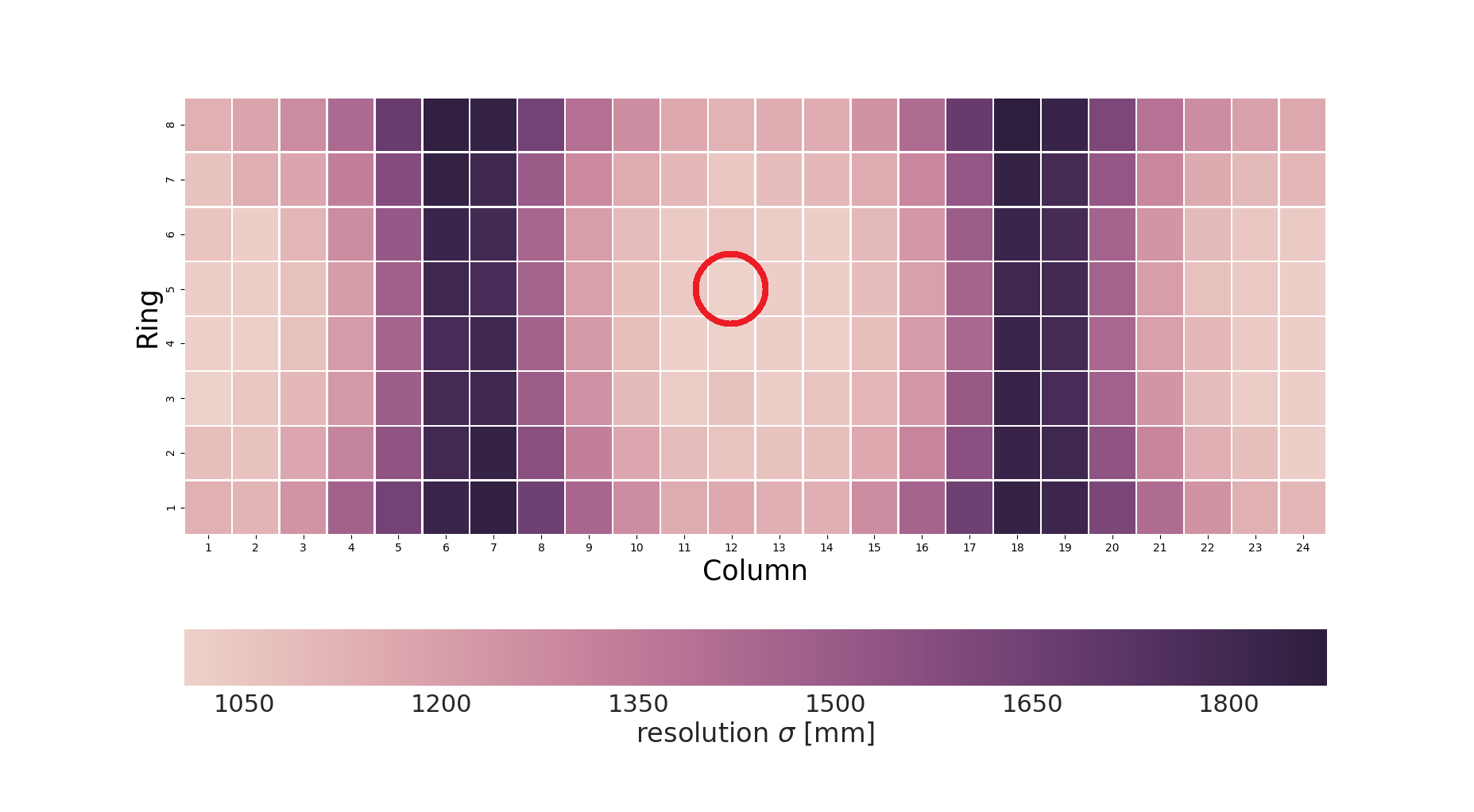}
		\label{fig_firstusub}
	}
	\subfigure[$\sigma_z(PMT_1)$] {
		\includegraphics[trim=110 0 110 0,clip,width=0.46\textwidth]{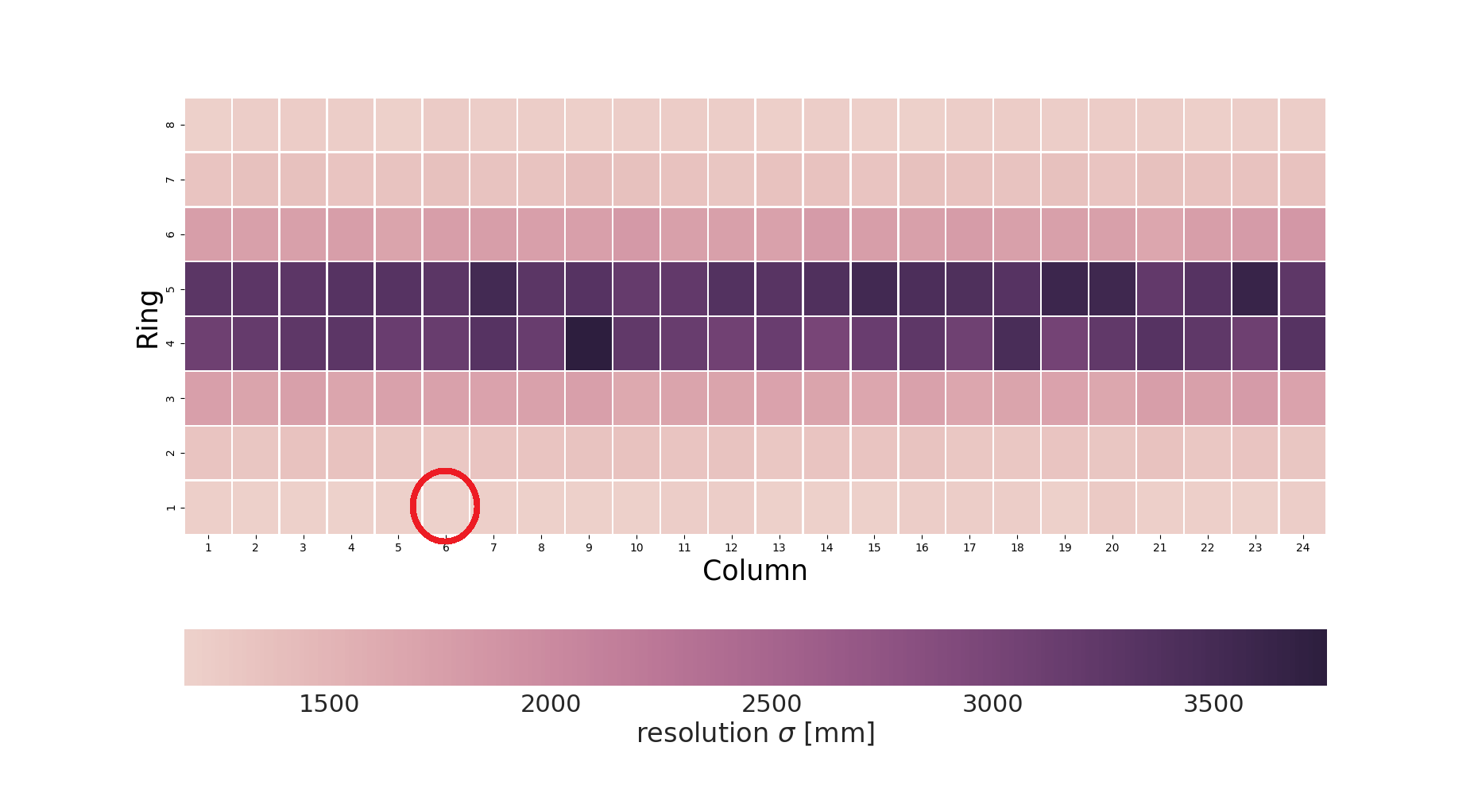}
		\label{fig_thirdusub}
	}
	\caption{Resolution $\sigma$ corresponding to using each individual PMT for (a) $x$ and (b) $z$, using the brute force search. In each heatmap, the circled PMT corresponds to the best resolution amongst all the 192 PMTs.}
	\label{greedy_firstsensor}
\end{figure}

The heatmap in Figures~\ref{greedy_firstsensor} show the resolutions as obtained from the residual distributions corresponding to using only one PMT for training and determining the vertex location of the antineutrino IBD interaction using the brute force search.
Specifically, the resolution pertaining to using each PMT is indicated by a value from the color scale. Clearly, some PMTs contains more information about the vertex position than others. The first most important PMT from the brute force search, i.e. $PMT_1^{brute*}$, is chosen as the one having the smallest color value in the heatmap. The variation in resolution for the $x$-direction by column is due to the use of $\sigma_x$ rather than $\sigma_r$. The reconstruction shows that the most important PMT is different for the $x$-direction and the $z$-direction. 
The heatmap pattern for the y-direction is similar to the x-direction, but with the dark region in the x-direction being the light region in the y-direction and vice versa, reflecting that the $x$ and $y$ depend on $\cos \phi$ and $\sin \phi$ respectively in a cylindrical coordinate system, i.e. a $\pi/2$ shift in difference between $x$ and $y$. 

\begin{figure}[!htbp]
	\centering
	\subfigure[$weight_x(PMT_1)$] {
		\includegraphics[trim=110 0 110 0,clip,width=0.46\textwidth]{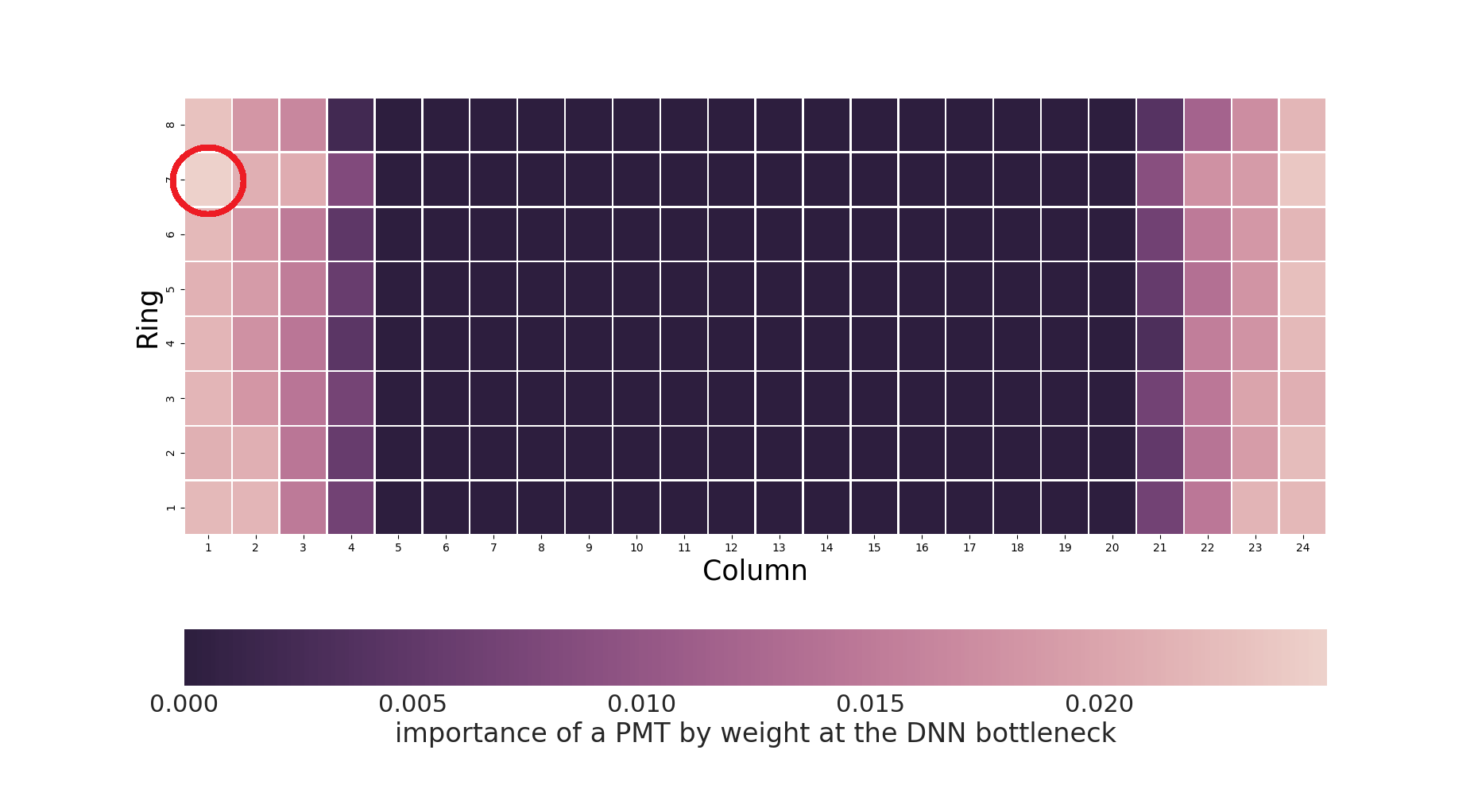}
		\label{fig_firstsub}
	}
	\subfigure[$weight_z(PMT_1)$] {
		\includegraphics[trim=110 0 110 0,clip,width=0.46\textwidth]{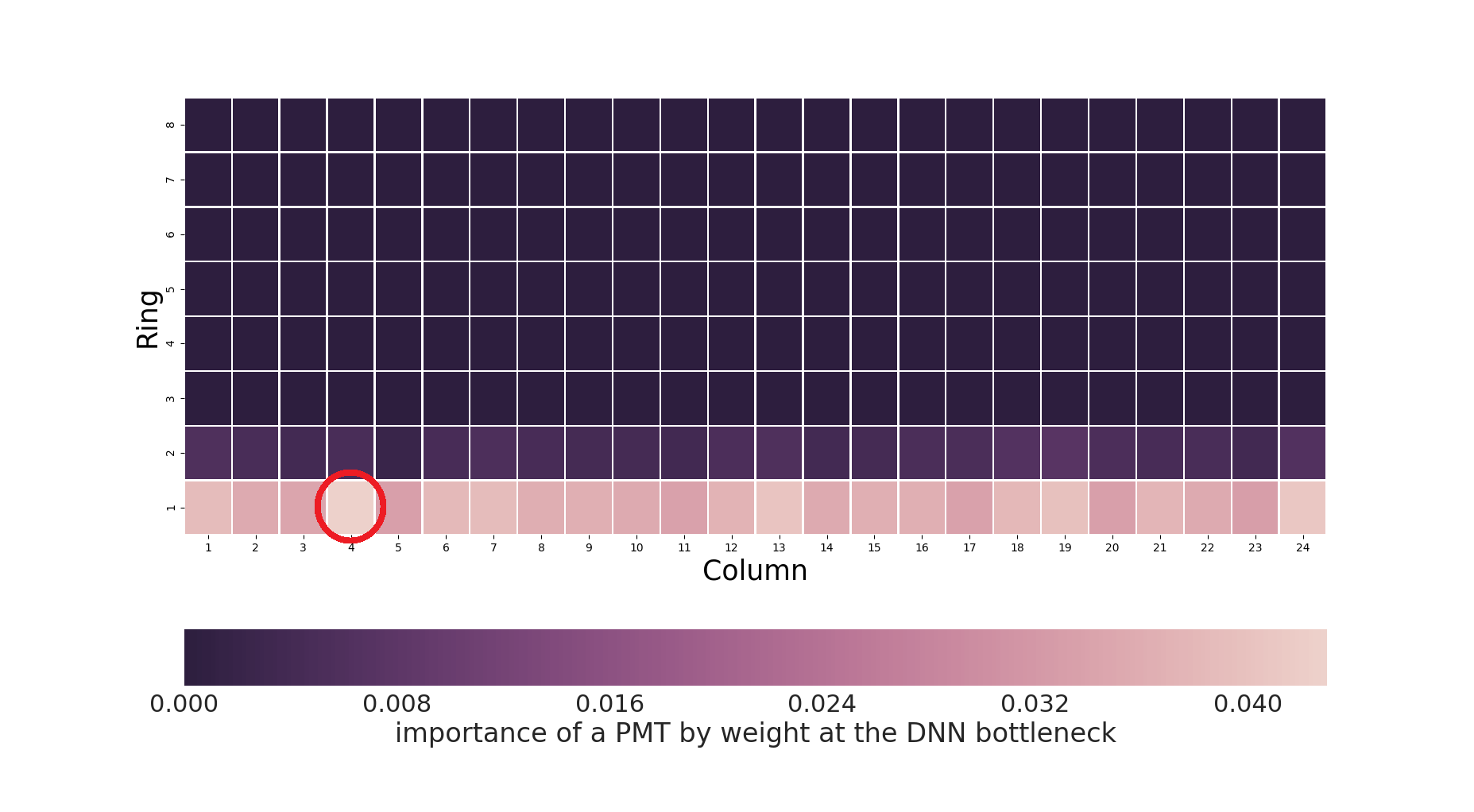}
		\label{fig_thirdsub}
	}
	\caption{Weights as given by the bottleneck neuron corresponding to each PMT for (a) $x$ and (b) $z$ using the bottleneck DNN. In each heatmap, the circled PMT corresponds to the largest weight in the heatmap amongst all the 192 PMTs, indicating that it is the most heavily used PMT in the DNN during the vertex reconstruction.}
	\label{bottleneck_firstsensor}
\end{figure}

\begin{figure}[!htbp]
	\centering
	\subfigure[$\sigma_x(PMT_2,PMT_1^{brute*})$] {
		\includegraphics[trim=110 0 110 0,clip,width=0.46\textwidth]{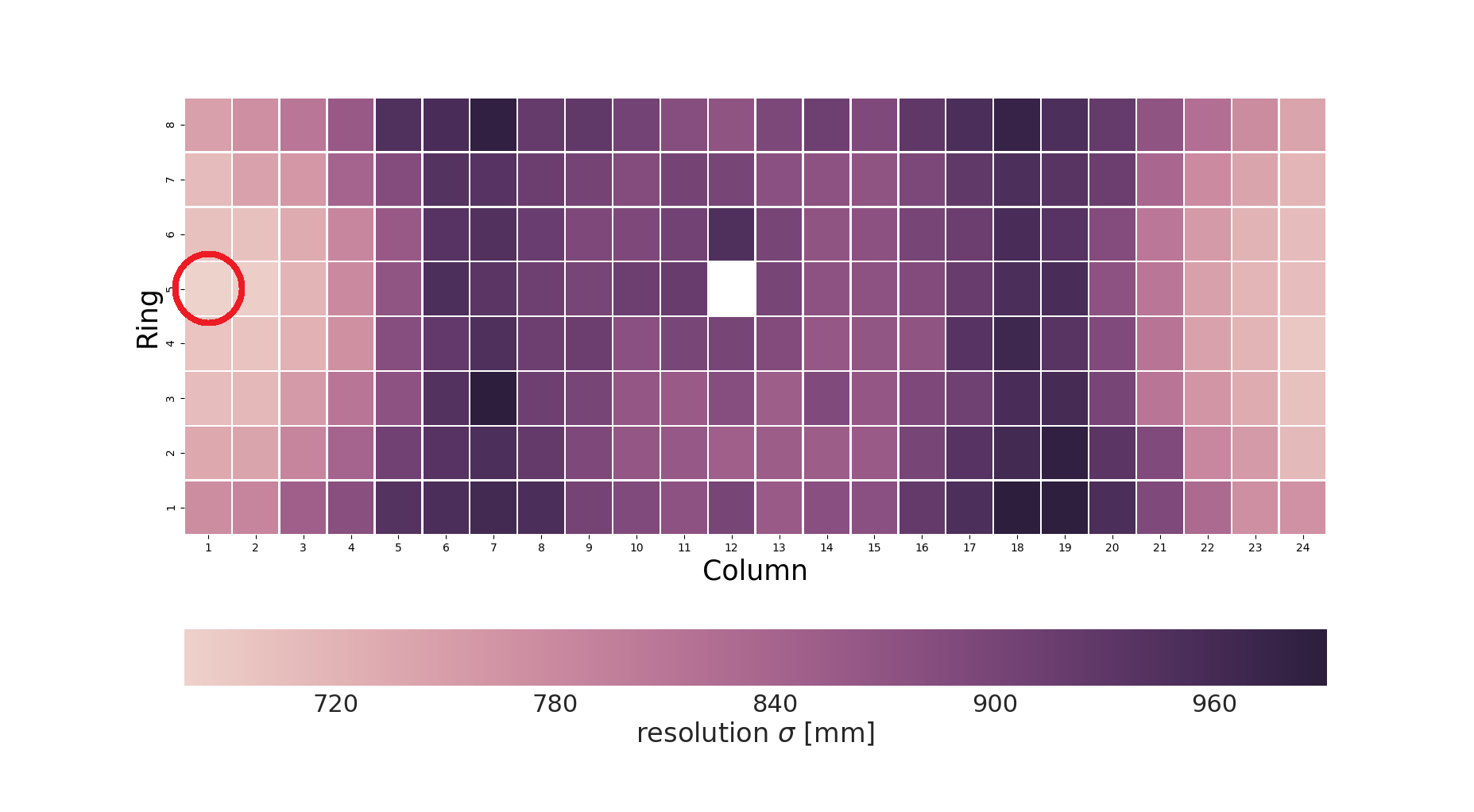}
		\label{fig_firstsub}
	}
	\subfigure[$\sigma_z(PMT_2,PMT_1^{brute*})$] {
		\includegraphics[trim=110 0 110 0,clip,width=0.46\textwidth]{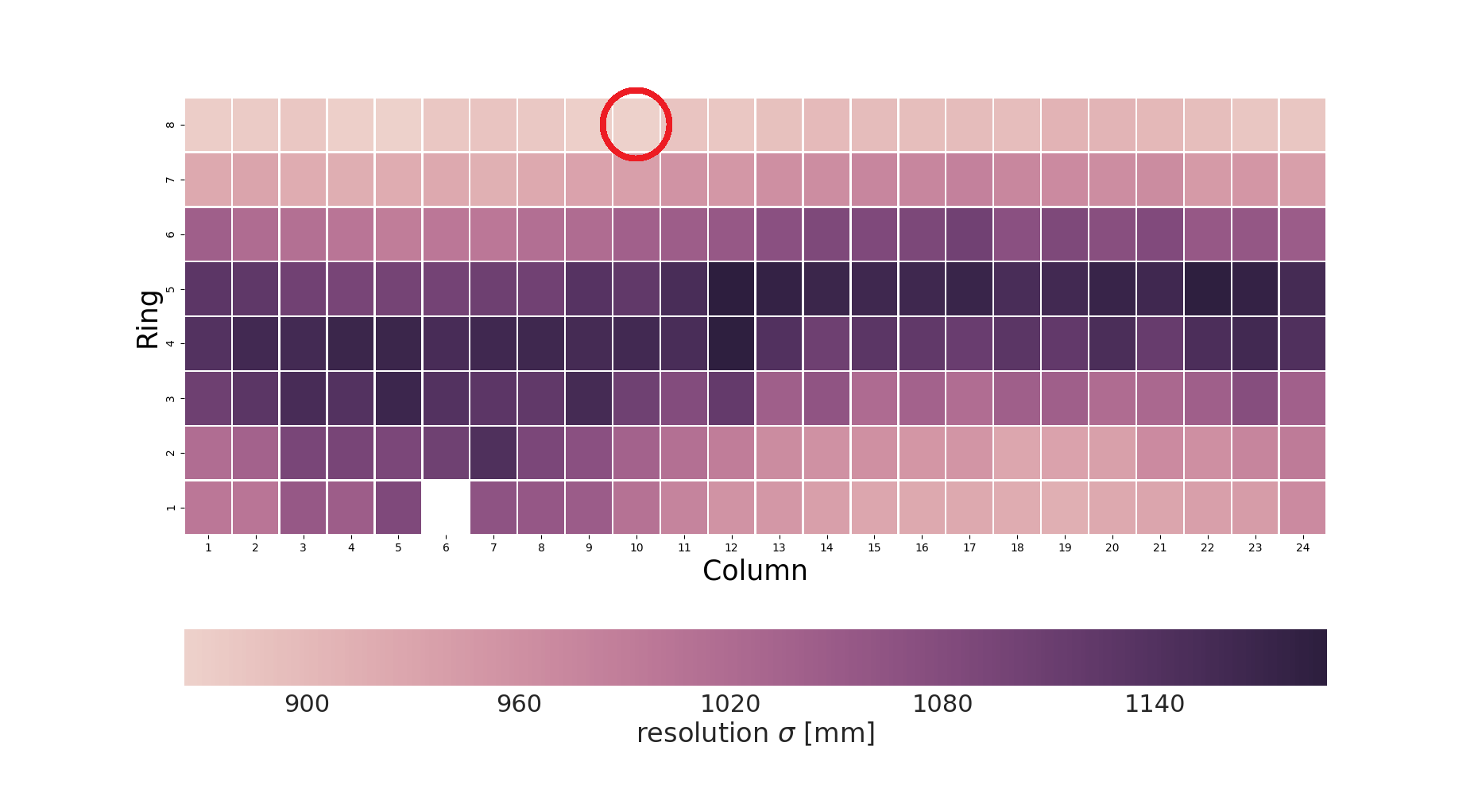}
		\label{fig_thirdosub}
	}
	\caption{Resolution $\sigma$ corresponding to using the charge information from the most important PMT as identified previously and a second candidate PMT for (a) $x$ and (b) $z$. The first PMT as found by the previous round, $PMT_1^{brute*}$ using the brute force search is whitened in each Figure. In each of the heatmap, the circled PMT, $PMT_2^*$ corresponds to the best resolution when considering the said PMT and the first important PMT found previously for the vertex reconstruction.  }
	\label{greedy_secondsensor}
\end{figure}

\begin{figure}[!htbp]
	\centering
	\subfigure[$weight_x(PMT_2 | PMT_1^{bneck*})$] {
		\includegraphics[trim=110 0 110 0 ,clip,width=.46\textwidth]{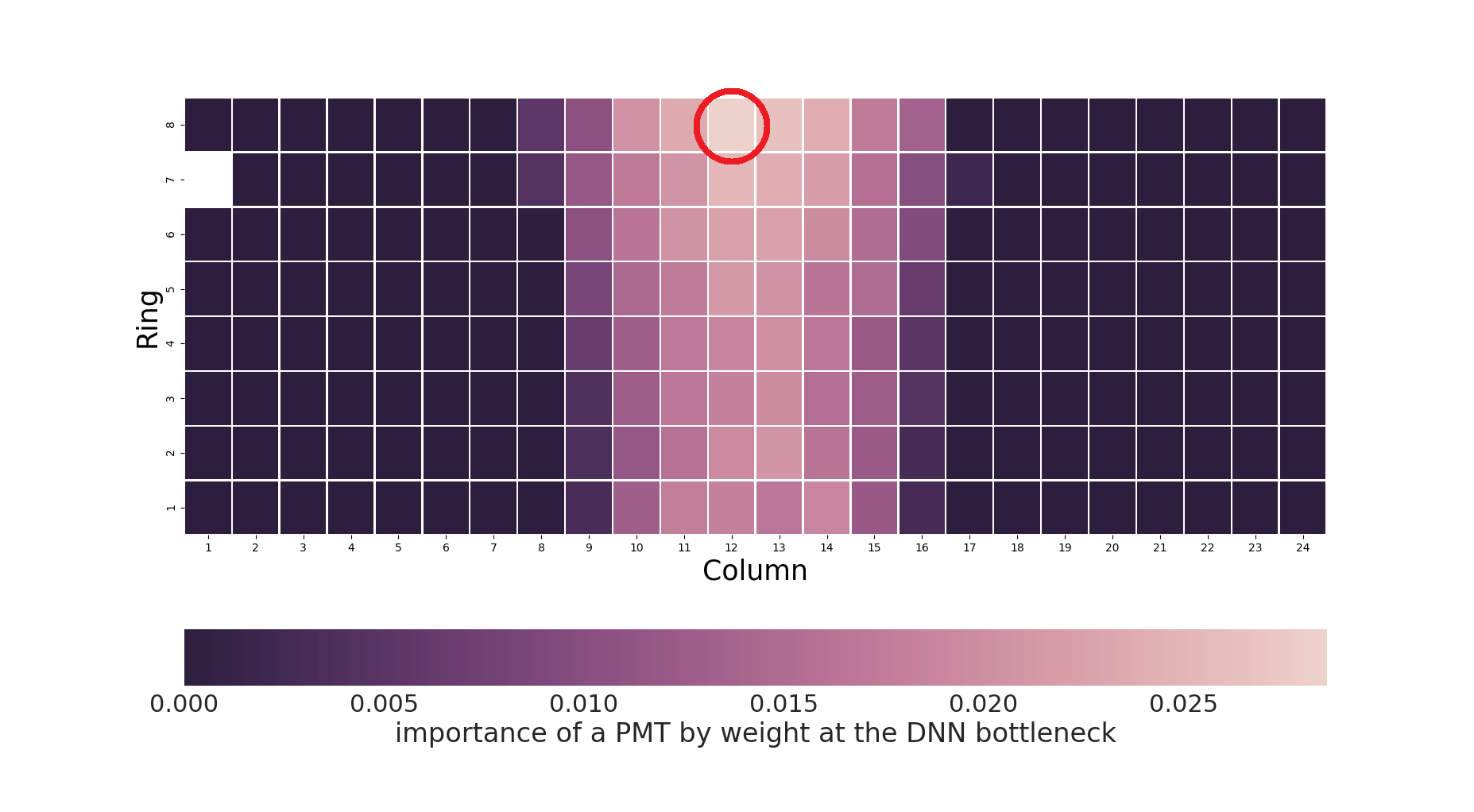}
		\label{fig_firstisub}
	}
	\subfigure[$weight_z(PMT_2 | PMT_1^{bneck*})$] {
		\includegraphics[trim=110 0 110 0,clip,width=0.46\textwidth]{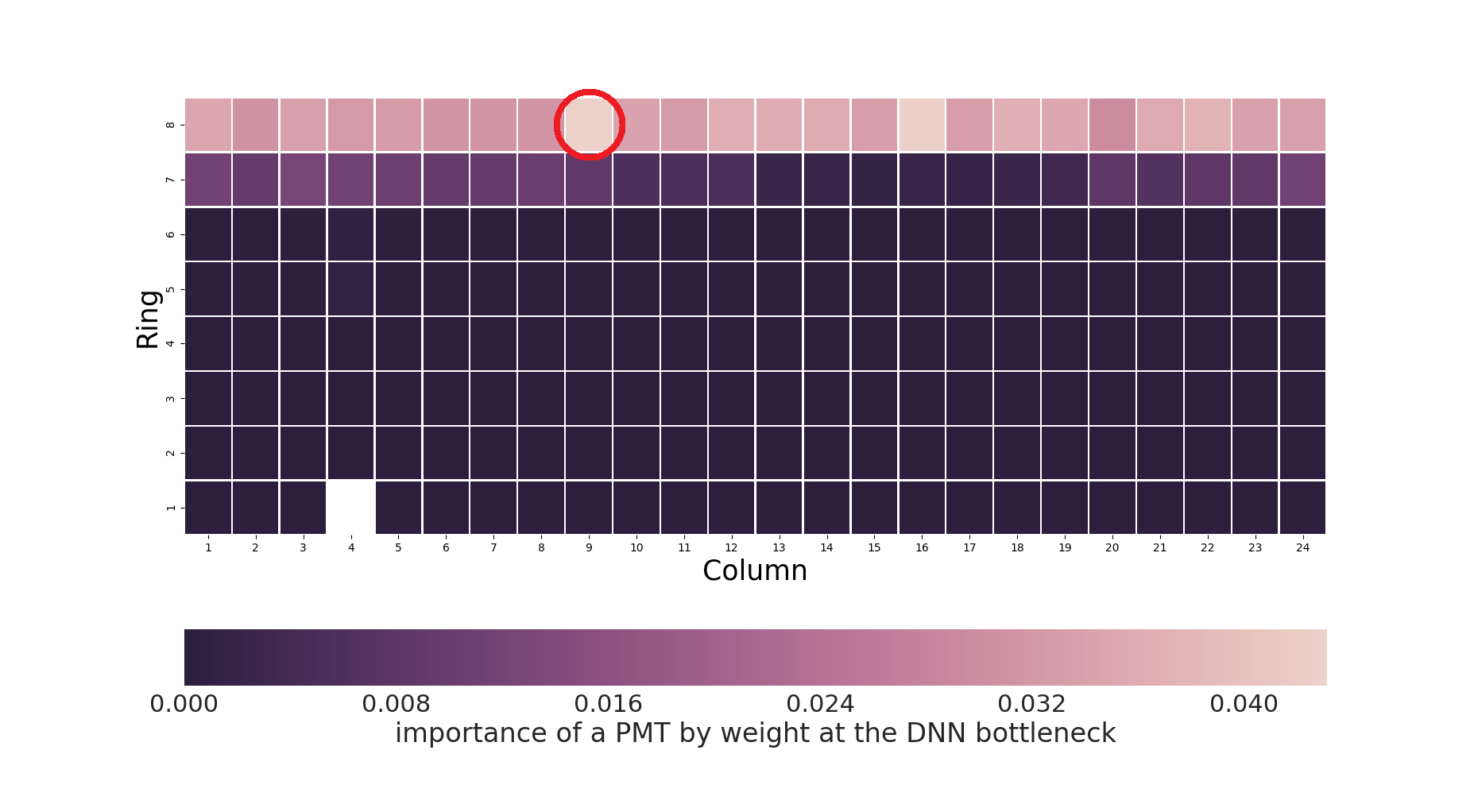}
		\label{fig_thirdisub}
	}
	\caption{Weights as given by the bottleneck neuron corresponding to using a second candidate PMT, and the most important PMT, $PMT_1^{bneck*}$ as identified previously for (a) $x$ and (b) $z$. $PMT_1^{bneck*}$ is whitened in each Figure. In each of the heatmap, the circled PMT, $PMT_2^*$ corresponds to the largest weight amongst the remaining 191 PMTs given that the first important PMT has been identified with the bottleneck DNN.  }
	\label{bottleneck_secondsensor}
\end{figure}

In Figure \ref{bottleneck_firstsensor}, the heatmap shows the weight corresponding to each PMT as obtained from the bottleneck neuron when searching for the first most important PMT. 
A higher weight indicates that the PMT has a larger impact on the vertex reconstruction. The PMT with the largest weight is identified as the most important PMT, $PMT_1^{bneck*}$.
Ideally, we would like to constrain the weights to be discrete at the bottleneck, i.e., to be either 0 or 1, where weight is 1 for the most important PMT and 0 for the rest during the training of the DNN. However, such a constraint is non-differentiable and non-continuous with respect to the loss function which would render DNN parameter optimization using gradient descent algorithms unfeasible. Comparing Figure \ref{greedy_firstsensor} and \ref{bottleneck_firstsensor}, the brute force and the bottleneck DNN have chosen different PMTs as their most important PMTs possibly due to degeneracies in the detector. For example, in the z-direction, the PMTs at the top and bottom rings should produce the same resolution. Our suspect is that during the training of the bottleneck DNN, some information sharing between a subset of PMTs, in which the DNN thinks their information values are similar, are unavoidable. Hence, the bottleneck neuron contains information from not one but a subset of PMTs, i.e. the importance by weights of each PMT could be partially "shared" amongst several PMTs. Further understanding of these are being conducted. 
Figures~\ref{greedy_secondsensor} and \ref{bottleneck_secondsensor} are the results as obtained from the brute force and bottleneck DNN approach respectively while searching for the second most important PMT after having found the first most important PMT.

\begin{figure}[!htbp]
	\centering
	\subfigure[$\sigma_x$ vs. number of PMTs] {
		\includegraphics[trim=110 0  100 0 , clip,width=0.46\textwidth]{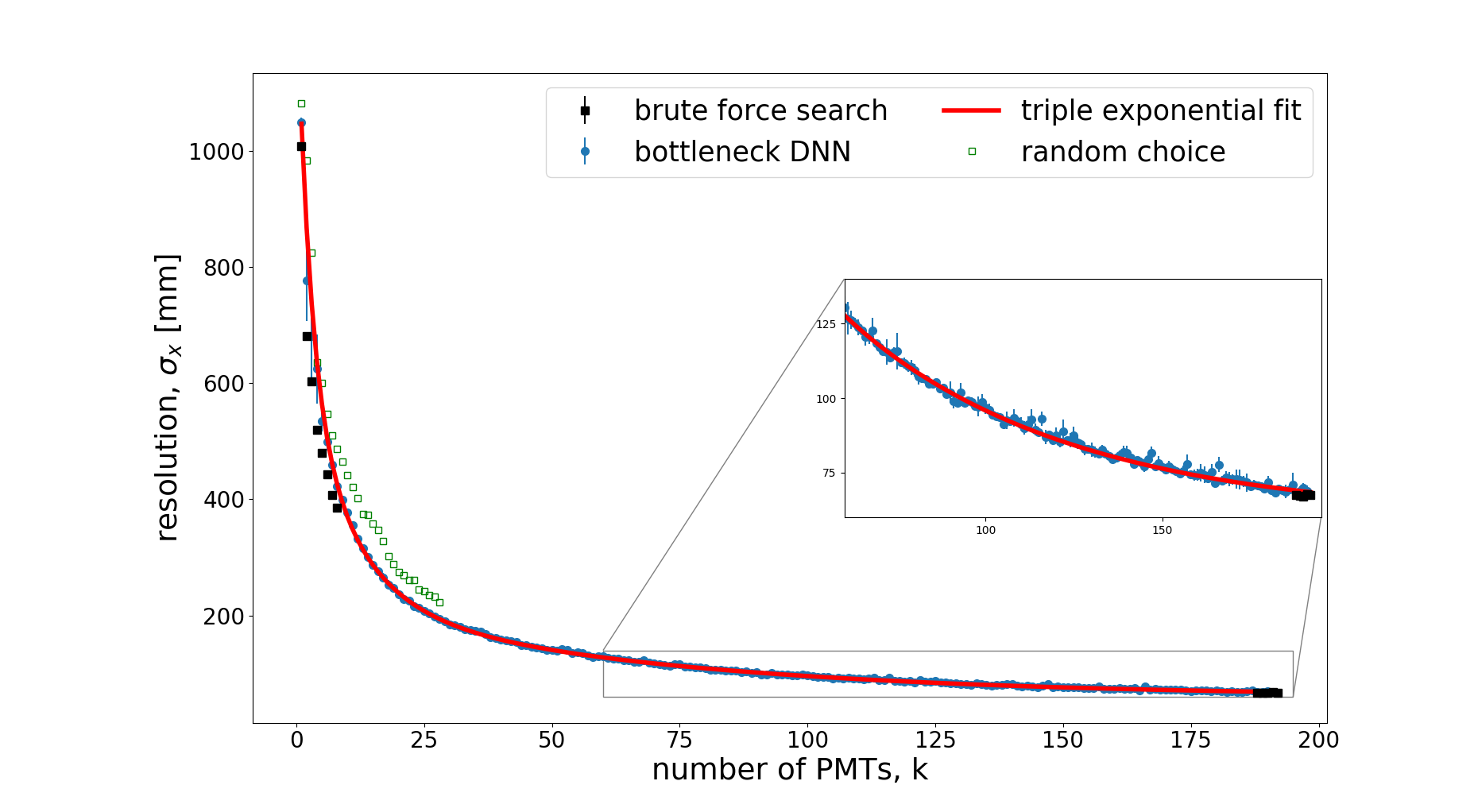}%_20170922}
		\label{fig_firstosub}
	}
	\subfigure[$\sigma_z$ vs. number of PMTs] {
		\includegraphics[trim=110 0 100 0 , clip,width=0.46\textwidth]{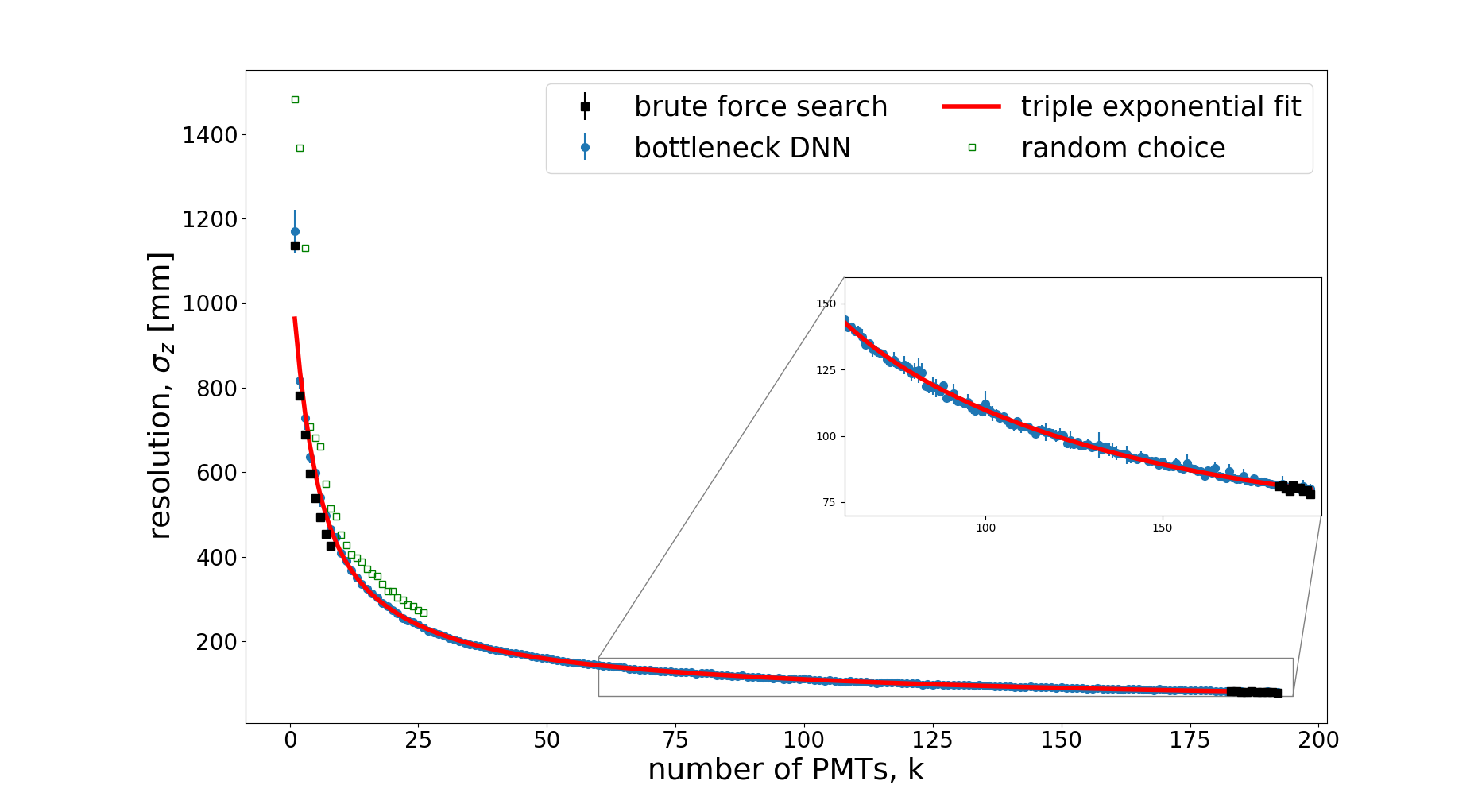}%_20170922}
		\label{fig_thirdsub}
	}
	\caption{Residual curves for $x$ (left Figure) and $z$ (right Figure) as a function of the number of PMTs used in the reconstruction. The results plotted as solid squares are those from the brute force search. An empirical fit to the bottleneck DNN results was done with a triple exponential. Resolutions from random choice of PMTs are shown as empty squares. A zoom-in on the residual curves is also shown in the Figures.}
	\label{residualcurve_xz}
\end{figure}

Figure \ref{residualcurve_xz} shows the residual curve for $x$ and $z$ as a function of the number of PMTs used in the reconstruction. Using an Nvidia Tesla P40 GPU, we estimated that it would take about 60 days to complete the entire residual curve with the brute force search, whereas it took about one day to complete with the bottleneck DNN. Resolutions from random choice of PMTs are also included in the Figures as a comparison to the result from the bottleneck DNN.
An empirical fit to the bottleneck DNN results is done with a triple exponential fit. It can be clearly seen that there is a diminishing return on the improvement in the vertex resolution when adding additional PMTs to an existing set of PMTs, which is an implication of the submodular \cite{Nemhauser} nature of the Gaussian standard deviation and its relationship to the information entropy, $H = \log \sigma + 0.5\log(2\pi e)$. Succinctly, the submodularity of the Gaussian standard deviation, i.e. the resolution in this case shows that there is less new information that could be gained from adding a new PMT to a larger set of already-chosen PMTs than to a smaller set.
As all the PMTs in the Daya Bay detector are of the same size and model, Figure \ref{residualcurve_xz} could also be interpreted as the residual curve being a function of the detector coverage.

\section{Summary}

In this work, we provide a fast approach using a deep neural network with a bottleneck neuron to uncover the effects of the number of photon sensors such as PMTs on the vertex resolutions in an antineutrino detector. The results have been compared with a random PMT search and a brute force search which yields the ideal result. Our inputs are the simulated charge information of the Daya Bay PMTs. The fast approach produces results close to those from the brute force search and fares much better than a random search.
We find that the vertex resolution of the event reconstruction at the Daya Bay is approximately a multi-exponentially decreasing function with respect to the number of PMTs and hence also, the coverage.
In future work, we envisage the possibility of incorporating the temporal information, i.e., the time of arrival of each photon in addition to the charge information to reconstruct the vertices. In addition, one could also study the size of the PMT needed alongside its installation location corresponding to the best event vertex reconstruction resolution. Also, a subsequent work from here would be the study of the effect of PMTs based on the event energy upon obtaining the vertex positions. Although studying the energy might need modifications to the deep network as the energy is a positive-definite quantity, the energy resolution is important when considering physics sensitivity, and thereby also impacting the design of future antineutrino detectors including JUNO. 

In order to use the bottleneck DNN approach for new detectors in designing phases, we suggest Monte Carlo simulations using various $N$. Then, one can obtain the $(P,N,k)$ surface in the hyperspace, where $P$ is some detector performance metric; $N$ and $k$ are as described in this work. Experimentalists can then decide on the $(P,N,k)$ working point for their detector in accordance with their construction budget and the desired detector performance.

\section{Conflicts of Interests}

The authors declare that there are no conflicts of interest regarding the publication of this paper.

\section*{Acknowledgements}

We thank Shen-Jian Chen and Zuo-Wei Liu for their computing facilities and helpful discussions. We would also like to thank Chao Zhang, Zhe Wang, Samuel Kohn, the Daya Bay ACC and the Collaboration for their time and comments. This work was supported by the National 973 Project Foundation of the Ministry of Science and Technology of China (Contract No. 2013CB834300).

\bibliography{paper}

\end{document}